\documentclass[aip,reprint,amsmath,amssymb,superscriptaddress,longbibliography,floatfix,10pt]{revtex4-1}

\usepackage[english]{babel}
\usepackage{bm,graphicx,braket,color,booktabs,fixmath}
\usepackage[hidelinks,colorlinks=true,citecolor=black,linkcolor=black,urlcolor=blue]{hyperref}

\begin{document}

\renewcommand{\vec}[1]{\mathbold{#1}}
\newcommand{\mat}[1]{\mathbold{#1}}

\title{Wave Physics as an Analog Recurrent Neural Network}

\author{Tyler W. Hughes}
\thanks{These authors contributed equally to this work.}
\affiliation{Department of Applied Physics, Stanford University, Stanford, CA 94305, USA}

\author{Ian A. D. Williamson}
\thanks{These authors contributed equally to this work.}
\affiliation{Department of Electrical Engineering, Stanford University, Stanford, CA 94305, USA}

\author{Momchil Minkov}
\affiliation{Department of Electrical Engineering, Stanford University, Stanford, CA 94305, USA}

\author{Shanhui Fan}
\email{shanhui@stanford.edu}
\affiliation{Department of Electrical Engineering, Stanford University, Stanford, CA 94305, USA}

\begin{abstract}
Analog machine learning hardware platforms promise to be faster and more energy-efficient than their digital counterparts.
Wave physics, as found in acoustics and optics, is a natural candidate for building analog processors for time-varying signals.
Here we identify a mapping between the dynamics of wave physics, and the computation in recurrent neural networks.
This mapping indicates that physical wave systems can be trained to learn complex features in temporal data, using standard training techniques for neural networks.
As a demonstration, we show that an inverse-designed inhomogeneous medium can perform vowel classification on raw audio signals as their waveforms scatter and propagate through it, achieving performance comparable to a standard digital implementation of a recurrent neural network. 
These findings pave the way for a new class of analog machine learning platforms, capable of fast and efficient processing of information in its native domain.
\end{abstract}

\maketitle

\section*{Introduction}

Recently, machine learning has had notable success in performing complex information processing tasks, such as computer vision \cite{russakovsky2015imagenet} and machine translation \cite{sutskever_sequence_2014}, which were intractable through traditional methods.
However, the computing requirements of these applications is increasing exponentially, motivating efforts to develop new, specialized hardware platforms for fast and efficient execution of machine learning models.
Among these are \textit{neuromorphic} hardware platforms \cite{shainline2017superconducting,tait_neuromorphic_2017,romera_vowel_2018}, with architectures that mimic the biological circuitry of the brain.
Furthermore, analog computing platforms, which use the natural evolution of a continuous physical system to perform calculations, are also emerging as important direction for implementation of machine learning \cite{shen_deep_2017,biamonte_quantum_2017,laporte2018numerical,lin2018all,khoram_stochastic_2018}.

In this work, we identify a mapping between the dynamics of wave-based physical phenomena, such as acoustics and optics, and the computation in a recurrent neural network (RNN).
RNNs are one of the most important machine learning models and have been widely used to perform tasks such as natural language processing \cite{yao2013recurrent} and time-series prediction \cite{husken_recurrent_2003,dorffner_neural_1996,connor_recurrent_1994}.
We show that wave-based physical systems can be trained to operate as an RNN, and as a result can \textit{passively} process signals and information in their native domain, without analog-to-digital conversion, which should result in a substantial gain in speed and a reduction in power consumption.
In this framework, rather than implementing circuits which deliberately route signals back to the input, the recurrence relationship occurs naturally in the time dynamics of the physics itself and the memory and capacity for information processing is provided by the waves as they propagate through space.

\section*{Results}

\subsection*{Equivalence between wave dynamics and a recurrent neural network}

\begin{figure*}
  \centering
  \includegraphics{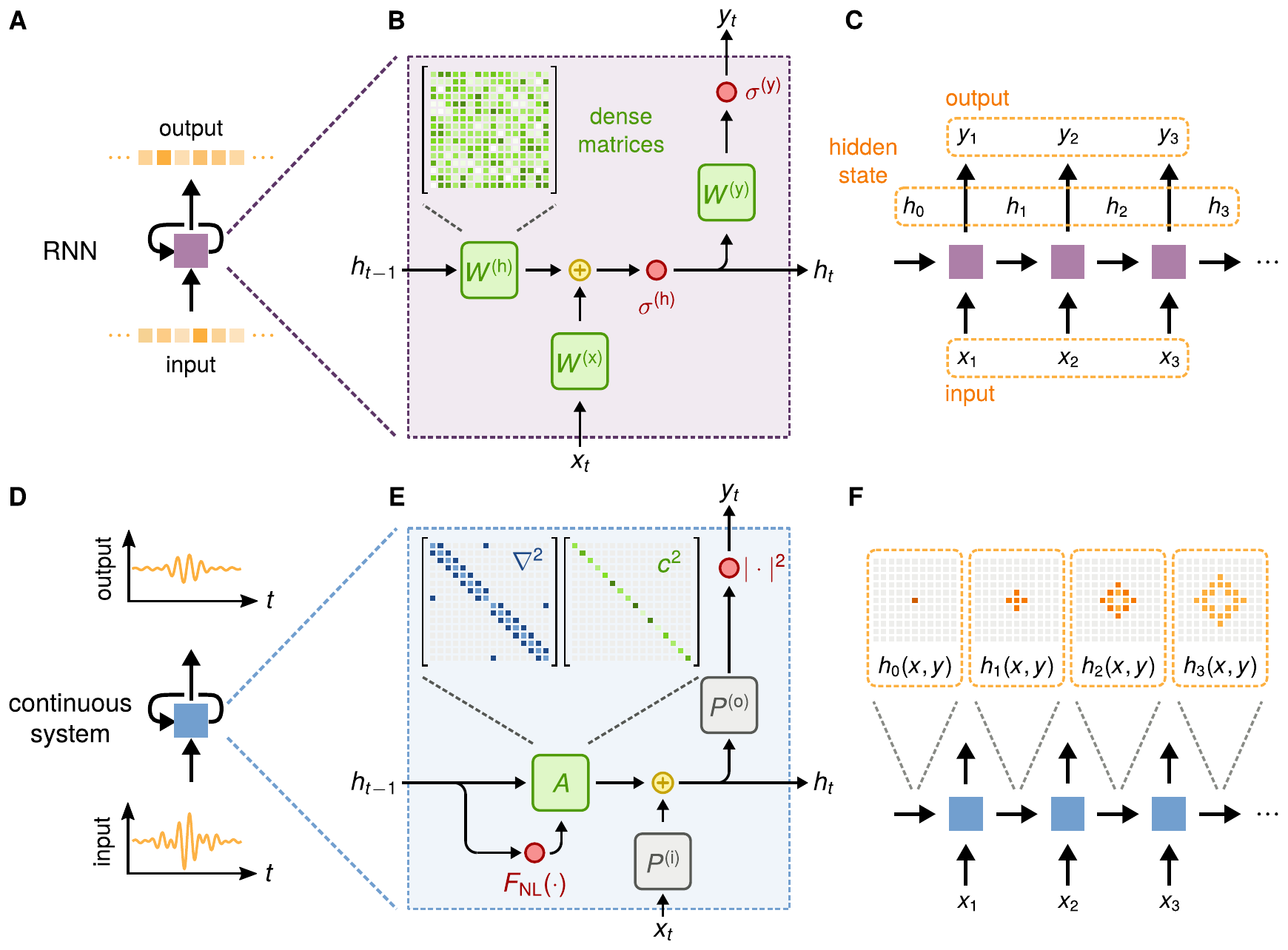}
  \caption{
  \textbf{Conceptual comparison of a standard recurrent neural network and a wave-based physcal system.}
  (\textbf{A})
  Diagram of a recurrent neural network (RNN) cell operating on a discrete input sequence and producing a discrete output sequence. 
  (\textbf{B})
  Internal components of the RNN cell, consisting of trainable dense matrices $\mat{W}^{(h)}$, $\mat{W}^{(x)}$, and $\mat{W}^{(y)}$. 
  Activation functions for the hidden state and output are represented by $\sigma^{(h)}$ and $\sigma^{(y)}$, respectively. 
  (\textbf{C}) 
  Diagram of the directed graph of the RNN cell. 
  (\textbf{D}) 
  Diagram of a recurrent representation of a continuous physical system operating on a continuous input sequence and producing a continuous output sequence. 
  (\textbf{E}) 
  Internal components of the recurrence relation for the wave equation when discretized using finite differences. 
  (\textbf{F}) 
  Diagram of the directed graph of discrete time steps of the continuous physical system and illustration of how a wave disturbance propagates within the domain.}
  \label{fig:fig1}
\end{figure*}

In this section we introduce the operation of an RNN and its connection to the dynamics of waves.
An RNN converts a sequence of inputs into a sequence of outputs by applying the same basic operation to each member of the input sequence in a step-by-step process (Fig \ref{fig:fig1}A). 
Memory of previous time steps is encoded into the RNN's \textit{hidden state}, which is updated at each step.
The hidden state allows the RNN to retain memory of past information and to learn temporal structure and long-range dependencies in data \cite{elman1990finding,jordan1997serial}.
At a given time step, $t$, the RNN operates on the current input vector in the sequence, $\vec{x}_t$, and the hidden state vector from the previous step, $\vec{h}_{t-1}$, to produce an output vector, $\vec{y}_t$, as well as an updated hidden state, $\vec{h}_t$. 
While many variations of RNNs exist, a common implementation \cite{Goodfellow-et-al-2016} is described by the following update equations
\begin{align}
    \vec{h}_t &= \sigma^{(h)}{\left(\mat{W}^{(h)} \cdot \vec{h}_{t-1} + \mat{W}^{(x)} \cdot \vec{x}_t \right)}
    \label{eq:eq1} \\
    \vec{y}_t &= \sigma^{(y)}{\left(\mat{W}^{(y)} \cdot \vec{h}_t\right)},
    \label{eq:eq2}
\end{align}
which are diagrammed in Fig. \ref{fig:fig1}B. 
The dense matrices defined by $\mat{W}^{(h)}$, $\mat{W}^{(x)}$, and $\mat{W}^{(y)}$ are optimized during training while $\sigma^{(h)}{\left(\cdot \right)}$ and $\sigma^{(y)}{\left(\cdot \right)}$ are nonlinear activation functions.
The operation prescribed by Eq. \ref{eq:eq1} and Eq. \ref{eq:eq2}, when applied to each element of an input sequence, can be described by the directed graph shown in Fig. \ref{fig:fig1}C.

We now discuss the connection between the dynamics in the RNN as described by Eq. \ref{eq:eq1} and Eq. \ref{eq:eq2}, and the dynamics of a wave-based physical system.
As an illustration, the dynamics of a scalar wave field distribution $u(x,y,z)$ are governed by the second-order partial differential equation (Fig. \ref{fig:fig1}D),
\begin{equation}
    \frac{\partial^2 u}{\partial t^2} - c^2 \cdot \nabla^2 u = f,
    \label{eq:eq3}
\end{equation}
where $\nabla^2 = \frac{\partial^2}{{\partial{x}}^2} + \frac{\partial^2}{{\partial{y}}^2} + \frac{\partial^2}{{\partial{z}}^2}$ is the Laplacian operator,
$c = c{\left(x,y,z\right)}$ is the spatial distribution of the wave speed, and
$f = f{\left(x,y,z,t\right)}$ is a source term. 
A finite-difference discretization of Eq. \ref{eq:eq3}, with a temporal step size of $\Delta{t}$, results in the recurrence relation,
\begin{equation}
        \frac{u_{t+1} - 2 u_t + u_{t-1}}{{\Delta{t}}^2} -  c^2\cdot \nabla^2 u_t = f_t.
        \label{eq:eq4}
\end{equation}
Here, the subscript, $t$, indicates the value of the scalar field at a fixed time step. 
The wave system's \textit{hidden state} is defined as the concatenation of the field distributions at the current and immediately preceding time steps, $\vec{h}_t \equiv [\vec{u}_t,~\vec{u}_{t-1}]^T$, where $\vec{u}_t$ and $\vec{u}_{t-1}$ are vectors given by the \textit{flattened} fields, $u_t$ and $u_{t-1}$, represented on a discretized grid over the spatial domain. 
Then, the update of the wave equation may be written as 
\begin{align}
\vec{h}_t &= \mat{A}{\left(\vec{h}_{t-1}\right)} \cdot \vec{h}_{t-1} + \mat{P}^{(\textrm{i})} \cdot \vec{x}_t \label{eq:eq5} \\
\vec{y}_t &= \left( \mat{P}^{(\textrm{o})} \cdot \vec{h}_t \right)^2, \label{eq:eq6}
\end{align}
where the sparse matrix, $\mat{A}$, describes the update of the wave fields $\vec{u}_t$ and $\vec{u}_{t-1}$ without a source (Fig. \ref{fig:fig1}E).  The derivation of Eq. \ref{eq:eq5} and \ref{eq:eq6} are given in supplementary materials section 1.

For sufficiently large field strengths, the dependence of $\mat{A}$ on $\vec{h}_{t-1}$ can be achieved through an intensity-dependent wave speed of the form $c = c_{\text{lin}} + {u_t}^2 \cdot c_{\text{nl}}$, where $c_{\text{nl}}$ is exhibited in regions of material with a nonlinear response.
In practice, this form of nonlinearity is encountered in a wide variety of wave physics, including shallow water waves \cite{ursell_long-wave_1953}, nonlinear optical materials via the Kerr effect \cite{boyd_nonlinear_2008}, and acoustically in bubbly fluids and soft materials \cite{rossing2007springer}.
Additional discussion and analysis of the nonlinear material responses is provided in supplementary materials section 2.
Like the $\sigma^{(y)}(\cdot)$ activation function in the standard RNN, a nonlinear relationship between the hidden state, $\vec{h}_t$, and the output, $\vec{y}_t$, of the wave equation is typical in wave physics when the output corresponds to a wave intensity measurement, as we assume here for Eq. \ref{eq:eq6}. 

Like the standard RNN, the connections between the hidden state and the input and output of the wave equation are also defined by linear operators, given by $\mat{P}^{(\textrm{i})}$ and $\mat{P}^{(\textrm{o})}$. 
These matrices define the injection and measuring points within the spatial domain.
Unlike the standard RNN, where the input and output matrices are dense, the input and output matrices of the wave equation are sparse because they are non-zero only at the location of injection and measurement points. 
Moreover, these matrices are unchanged by the training process.
Additional information on the construction of these matrices is given in supplementary materials section 3.

Most importantly, the trainable free parameter of the wave equation is the distribution of the wave speed, $c{\left(x,y,z\right)}$. 
In practical terms, this corresponds to the physical configuration and layout of materials within the domain.
Thus, when modeled numerically in discrete time (Fig. \ref{fig:fig1}E), the wave equation defines an operation which maps into that of an RNN (Fig. \ref{fig:fig1}B).
Similarly to the RNN, the full time dynamics of the wave equation may be represented as a directed graph, but in contrast, the nearest-neighbor coupling enforced by the Laplacian operator leads to information propagating through the hidden state with a finite velocity (Fig. \ref{fig:fig1}F).

\subsection*{Training a physical system to classify vowels}

\begin{figure*}
  \centering
  \includegraphics{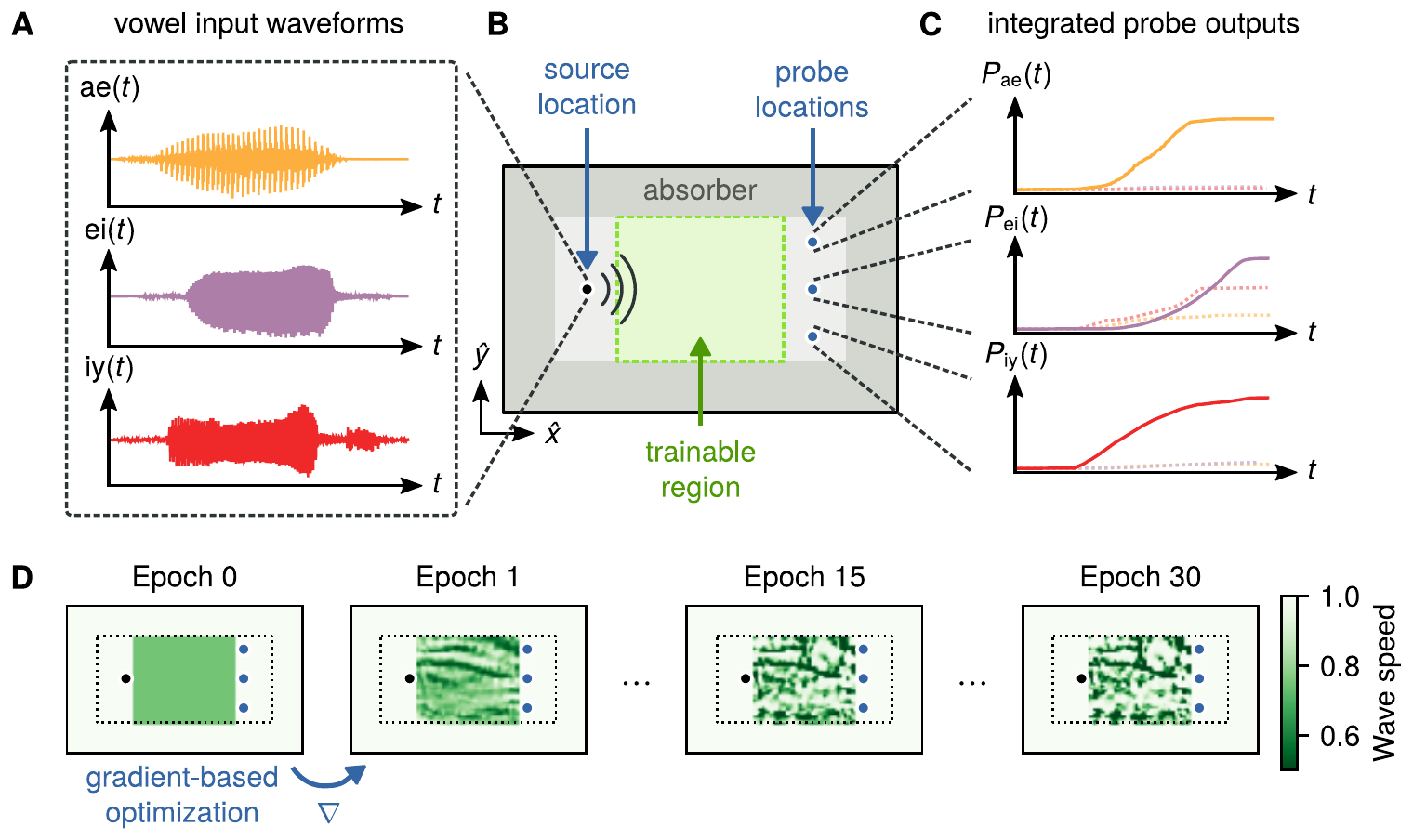}
  \caption{
  \textbf{Schematic of the vowel recognition setup and the training procedure.}
  (\textbf{A})
  Raw audio waveforms of spoken vowel samples from three classes.
  (\textbf{B})
  Layout of the vowel recognition system. Vowel samples are independently injected at the source, located at the left of the domain, and propagate through the center region, indicated in green, where a material distribution is optimized during training. The dark gray region represents an absorbing boundary layer.
  (\textbf{C})
  For classification, the time-integrated power at each probe is measured and normalized to be interpreted as a probability distribution over the vowel classes.
  (\textbf{D})
  Using automatic differentiation, the gradient of the loss function with respect to the density of material in the green region is computed. The material density is updated iteratively, using gradient-based stochastic optimization techniques, until convergence.
  }
  \label{fig:fig2}
\end{figure*}

We now demonstrate how the dynamics of the wave equation can be trained to classify vowels through the construction of an inhomogeneous material distribution. 
For this task, we utilize a dataset consisting of 930 raw audio recordings of 10 vowel classes from 45 different male speakers and 48 different female speakers \cite{hillenbrand_acoustic_1995}. 
For our learning task, we select a subset of 279 recordings corresponding to three vowel classes, represented by the vowel sounds \textit{ae}, \textit{ei}, and \textit{iy}, as contained in the words h\textit{a}d, h\textit{aye}d, and h\textit{ee}d, respectively (Fig. \ref{fig:fig2}A).

The physical layout of the vowel recognition system consists of a two-dimensional domain in the $x$-$y$ plane, infinitely extended along the $z$-direction (Fig. \ref{fig:fig2}B). 
The audio waveform of each vowel, represented by $\vec{x}^{(i)}$, is injected by a source at a single grid cell on the left side of the domain, emitting waveforms which propagate through a central region with a trainable distribution of the wave speed, indicated by the light green region in Fig. \ref{fig:fig2}B. 
Three probe points are defined on the right hand side of this region, each assigned to one of the three vowel classes. 
To determine the system's output, $\vec{y}^{(i)}$, the time-integrated power at each probe is measured (Fig. \ref{fig:fig2}C). 
After the simulation evolves for the full duration of the vowel recording, this integral gives a non-negative vector of length 3, which is then normalized by its sum and interpreted as the system's predicted probability distribution over the vowel classes. 
An absorbing boundary region, represented by the dark gray region in Fig. \ref{fig:fig2}B, is included in the simulation to prevent energy from building up inside the computational domain. 
The derivation of the discretized wave equation with a damping term is given in supplementary materials section 1.

For the purposes of our numerical demonstration, we consider binarized systems consisting of two materials: a background material with a normalized wave speed $c_0 = 1.0$, and a second material with $c_1 = 0.5$.
We assume that the second material has a nonlinear parameter, $c_{\text{nl}} = -30$, while the background material has a linear response.
In practice, the wave speeds would be modified to correspond to different materials being used.
For example, in an acoustic setting the material distribution could consist of air, where the sound speed is 331 m/s, and porous silicone rubber, where the sound speed is 150 m/s \cite{ba_soft_2017}.
The initial distribution of the wave speed consists of a uniform region of material with a speed which is midway between those of the two materials (Fig. \ref{fig:fig2}D).
This choice of starting structure allows for the optimizer to shift the density of each pixel towards either one of the two materials to produce a binarized structure consisting of only those two materials. 
To train the system, we perform back-propagation through the model of the wave equation to compute the gradient of the cross entropy loss function of the measured outputs with respect to the density of material in each pixel of the trainable region.
Interestingly, this approach is mathematically equivalent to the \textit{adjoint method} \cite{hughes_training_2018}, which is widely used for inverse design \cite{molesky2018inverse,hughes_training_2018,elesin_design_2012}.
Then, we use this gradient information update the material density using the the Adam optimization algorithm \cite{kingma_adam_2014}, repeating until convergence on a final structure (Fig. \ref{fig:fig2}D).

\begin{figure*}
  \centering
  \includegraphics{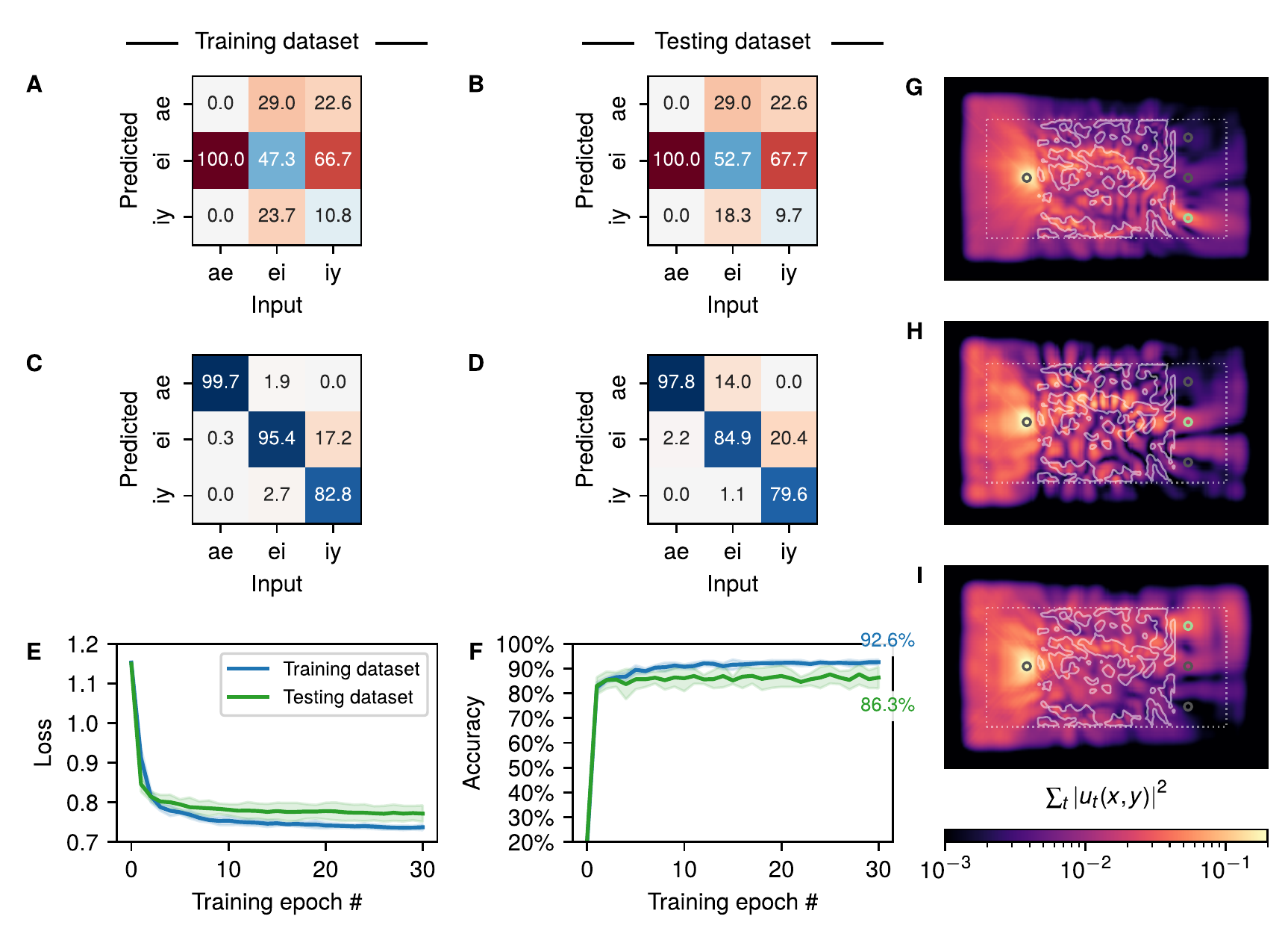}
  \caption{\textbf{Vowel recognition training results.}
  Confusion matrix over the training and testing datasets for the initial structure (\textbf{A}),(\textbf{B}) and final structure (\textbf{C}),(\textbf{D}), indicating the percentage or correct (diagonal) and incorrect (off-diagonal). 
  Cross validated training results showing the mean (sold line) and standard deviation (shaded region) of the (\textbf{E}) cross entropy loss and (\textbf{F}) prediction accuracy over 30 training epochs and 5 folds of the dataset, which consists of a total of 279 total vowel samples of male and female speakers.
  (\textbf{G})-(\textbf{I}) The time-integrated intensity distribution for a randomly selected input (\textbf{G}) \textit{ae} vowel, (\textbf{H}) \textit{ei} vowel, and (\textbf{I}) \textit{iy} vowel.}
  \label{fig:fig3}
\end{figure*}

The confusion matrices over the training and testing sets for the starting structure are shown in Fig. \ref{fig:fig3}A and Fig. \ref{fig:fig3}B, averaged over five cross-validated training runs.
Here, the confusion matrix defines the percentage of correctly predicted vowels along its diagonal entries and the percentage of incorrectly predicted vowels for each class in its off-diagonal entries.
Clearly the starting structure can not perform the recognition task.
Fig. \ref{fig:fig3}C and Fig. \ref{fig:fig3}D show the final confusion matrices after optimization for the testing and training sets, averaged over five cross validated training runs.
The trained confusion matrices are diagonally dominant, indicating that the structure can indeed perform vowel recognition. More information on the training procedure and numerical modeling is provided in the materials and methods section.

Fig. \ref{fig:fig3}E and Fig. \ref{fig:fig3}F show the cross entropy loss value and the prediction accuracy, respectively, as a function of the training epoch over the testing and training datasets, where the solid line indicates the mean and the shaded region corresponds to the standard deviation over the cross-validated training runs.
Interestingly, we observe that the first epoch results in the largest reduction of the loss function and the largest gain in prediction accuracy.
From Fig. \ref{fig:fig3}F we see that the system obtains a mean accuracy of 92.6\% $\pm$ 1.1\% over the training dataset and a mean accuracy of 86.3\% $\pm$ 4.3\% over the testing dataset.
From Fig. \ref{fig:fig3}C and Fig. \ref{fig:fig3}D we observe that the system attains near perfect prediction performance on the \textit{ae} vowel and is able to differentiate the \textit{iy} vowel from the \textit{ei} vowel, but with less accuracy, especially in unseen samples from the testing dataset. 
Fig. \ref{fig:fig3}G, Fig. \ref{fig:fig3}H, and Fig. \ref{fig:fig3}I show the distribution of the integrated field intensity, $\sum_t{ {u_t}^2}$, when a representative sample from each vowel class is injected into the trained structure. 
We thus provide visual confirmation that the optimization procedure produces a structure which routes the majority of the signal energy to the correct probe. 
As a performance benchmark, a conventional RNN was trained on the same task, achieving comparable classification accuracy to that of the wave equation. However, a larger number of free parameters was required.
Additionally, we observed that a comparable classification accuracy was obtained when training a linear wave equation.
More details on the performance are provided in supplementary materials section 4.

\section*{Discussion}

The wave-based RNN presented here has a number of favorable qualities that make it a promising candidate for processing temporally-encoded information.
Unlike the standard RNN, the update of the wave equation from one time step to the next enforces a nearest-neighbor coupling between elements of the hidden state through the Laplacian operator, which is represented by the sparse matrix in Fig. \ref{fig:fig1}E. 
This nearest neighbor coupling is a direct consequence of the fact that the wave equation is a hyperbolic partial differential equation in which information propagates with a finite velocity. 
Thus, the size of the analog RNN's hidden state, and therefore its memory capacity, is directly determined by the size of the propagation medium. 
Additionally, unlike the conventional RNN, the wave equation enforces an energy conservation constraint, preventing unbounded growth of the norm of the hidden state and the output signal.
In contrast, the unconstrained dense matrices defining the update relationship of the standard RNN lead to vanishing and exploding gradients, which can pose a major challenge for training traditional RNNs \cite{jing2017tunable}.

We have shown that the dynamics of the wave equation are conceptually equivalent to those of a recurrent neural network.
This conceptual connection opens up the opportunity for a new class of analog hardware platform, in which evolving time dynamics play a major role in both the physics and the dataset.
While we have focused on a the most general example of wave dynamics, characterized by a scalar wave equation, our results can be readily extended to other wave-like physics.
Such an approach of using physics to perform computation \cite{silva_performing_2014,hermans_trainable_2015,guo_photonic_2018,lin2018all,kwon_nonlocal_2018,estakhri_inversedesigned_2019} may inspire a new platform for analog machine learning devices, with the potential to perform computation far more naturally and efficiently than their digital counterparts.
The generality of the approach further suggests that many physical systems may be attractive candidates for performing RNN-like computations on dynamic signals, such as those in optics, acoustics, or seismics.

\section*{Materials and Methods}

\subsection*{Training method for vowel classification}

The procedure for training the vowel recognition system is as follows. 
First, each vowel waveform is downsampled from its original recording, with a 16 kHz sampling rate, to a sampling rate of 10 kHz.
Next, the entire dataset of (3 classes) $\times$ (45 males + 48 females) = 279 vowel samples is divided into 5 groups of approximately equal size.

Cross validated training is performed with 4 out of the 5 sample groups forming a training set and 1 out of the 5 sample groups forming a testing set.
Independent training runs are performed with each of the 5 groups serving as the testing set, and the metrics are averaged over all training runs. 
Each training run is performed for 30 epochs using the Adam optimization algorithm \cite{kingma_adam_2014} with a learning rate of 0.0004. 
During each epoch, every sample vowel sequence from the training set is windowed to a length of 1000, taken from the center of the sequence.
This limits the computational cost of the training procedure by reducing the length of the time through which gradients must be tracked.

All windowed samples from the training set are run through the simulation in batches of 9 and the categorical cross entropy loss is computed between the output probe probability distribution and the correct one-hot vector for each vowel sample.
To encourage the optimizer to produce a binarized distribution of the wave speed with relatively large feature sizes, the optimizer minimizes this loss function with respect to a material density distribution, $\rho{\left(x,y\right)}$ within a central region of the simulation domain, indicated by the green region in Fig. \ref{fig:fig2}B. 
The distribution of the wave speed, $c{\left(x,y\right)}$, is computed by first applying a low-pass spatial filter and then a projection operation to the density distribution. 
The details of this process are described in supplementary materials section 5.
Fig. \ref{fig:fig2}D illustrates the optimization process over several epochs, during which, the wave velocity distribution converges to a final structure.
At the end of each epoch, the classification accuracy is computed over both the testing and training set. 
Unlike the training set, the full length of each vowel sample from the testing set is used.

\begin{figure}
  \centering
  \includegraphics[width=\columnwidth]{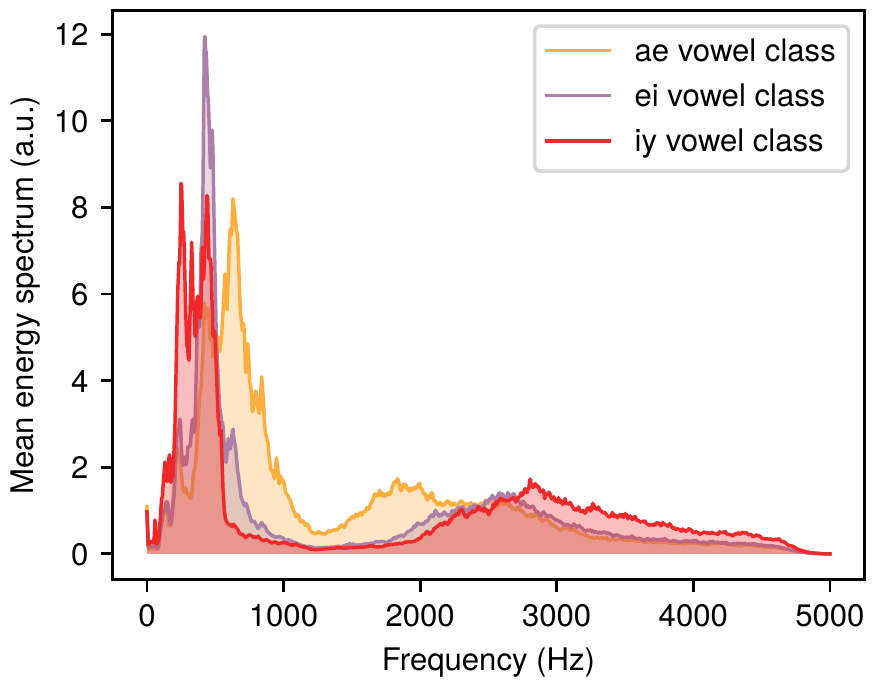}
  \caption{\textbf{Frequency content of the vowel classes.} The plotted quantity is the mean energy spectrum for the \textit{ae}, \textit{ei}, and \textit{iy} vowel classes.}
  \label{fig:fig4}
\end{figure}

The mean energy spectrum of the three vowel classes after downsampling to 10 kHz is shown in Fig. \ref{fig:fig4}.
We observe that the majority of the energy for all vowel classes is below 1 kHz and that there is strong overlap between the mean peak energy of the \textit{ei} and \textit{iy} vowel classes.
Moreover, the mean peak energy of the \textit{ae} vowel class is very close to the peak energy of the other two vowels.
Therefore, the vowel recognition task learned by the system is non-trivial.

\subsection*{Numerical modeling}

Numerical modeling and simulation of the wave equation physics was performed using a custom package written in Python.
The software was developed on top of the popular machine learning library, \texttt{pytorch} \cite{paszke2017automatic}, to compute the gradients of the loss function with respect to the material distribution via reverse-mode automatic differentiation.
In the context of inverse design in the fields of physics and engineering, this method of gradient computation is commonly referred to as the adjoint variable method and has a computational cost of performing one additional simulation.
We note that related approaches to numerical modeling using machine learning frameworks have been proposed previously for full-wave inversion of seismic datasets \cite{richardson_seismic_2018}.
The code for performing numerical simulations and training of the wave equation, as well as generating the figures presented in this paper, may be found online at \url{http://www.github.com/fancompute/wavetorch/}.

\section*{Acknowledgements}

\textbf{Funding:} This work was supported by a Vannevar Bush Faculty Fellowship from the U.S. Department of Defense (N00014-17-1-3030), by the Gordon and Betty Moore Foundation (GBMF4744), by a MURI grant from the U.S. Air Force Office of Scientific Research (FA9550-17-1-0002), and by the Swiss National Science Foundation (P300P2\_177721).
\textbf{Author contributions:} T.W.H and I.A.D.W. contributed equally to this work. T.W.H. conceived the idea and developed it with I.A.D.W., with input from M.M. The software for performing numerical simulations andtraining of the wave equation was developed by I.A.D.W. with input from T.W.H. and M.M. The numerical model for the standard RNN was developed and trained by M.M. S.F. supervised the project. All authors contributed to analyzing the results and writing the manuscript.
\textbf{Competing interests:} All authors have jointly filed for a provisional patent on the idea. The authors declare no other competing interests. 
\textbf{Data and materials availability:} All data needed to evaluate the conclusions in the paper are present in the paper and/or the Supplementary Materials. Additional data related to this paper may be requested from the authors. The code for performing numerical simulations and training of the wave equation, as well as generating the figures presented in this paper are available online at \url{http://www.github.com/fancompute/wavetorch/}.

\onecolumngrid
\clearpage

\section*{Supplementary Materials}

\renewcommand{\thesection}{S\arabic{section}}
\renewcommand{\thesubsection}{S\arabic{subsection}}
\setcounter{section}{0}
\renewcommand{\thefigure}{S\arabic{figure}}
\setcounter{figure}{0}
\renewcommand{\thetable}{S\arabic{table}} 
\setcounter{table}{0}
\renewcommand{\theequation}{S\arabic{equation}} 
\setcounter{equation}{0}
\renewcommand{\thepage}{S\arabic{page}} 
\setcounter{page}{1}

\subsection{Derivation of the wave equation update relationship \label{sec:scalar_wave}}

In the main text, we specified that the dynamics of the scalar field distribution, $u = u{\left(x,y,z,t\right)}$, are governed by the wave equation
\begin{equation}
    \frac{\partial^2{u}}{{\partial{t}}^2} - c^2 \cdot \nabla^2 u = f,
    \label{eq:sup_UndampedScalarWave}
\end{equation}
where $\nabla^2 = \frac{\partial^2}{{\partial{x}}^2} + \frac{\partial^2}{{\partial{y}}^2} + \frac{\partial^2}{{\partial{z}}^2}$ is the Laplacian operator.
$c = c{\left(x,y,z\right)}$ is the spatial distribution of the wave speed and $f = f{\left(x,y,z,t\right)}$ is a source term.  
As discussed in the main text, nonlinear materials have a wave speed which depends on the wave amplitude. 
Equation \ref{eq:sup_UndampedScalarWave} can be discretized in time using centered finite differences with a temporal step size of $\Delta{t}$, after which it becomes
\begin{equation}
        \frac{u_{t+1} - 2 u_t + u_{t-1}}{{\Delta{t}}^2} -  c^2\cdot \nabla^2 u_t = f_t.
        \label{eq:sup_finite_diff}
\end{equation}
Here, the subscript $t$ is used to indicate the value of a scalar field at a given time step. To connect Eq. \ref{eq:sup_finite_diff} to the RNN update equations from Eq. \ref{eq:eq1} and \ref{eq:eq2}, we exress this in matrix form as
\begin{equation}
    \begin{bmatrix}
    u_{t+1} \\ u_t
    \end{bmatrix}
    = 
    \begin{bmatrix}
    2 + \Delta t^2\cdot c^2 \cdot \nabla^2 
    & -1  \\
    1 & 0
    \end{bmatrix}
    \cdot
    \begin{bmatrix}
    u_{t} \\ u_{t-1}
    \end{bmatrix}    
    +
    \Delta{t}^2 \cdot \begin{bmatrix}
     f_{t} \\ 0
    \end{bmatrix}.
    \label{eq:matrix_form}
\end{equation}
Then, the update equation for the wave equation defined by Eq. \ref{eq:matrix_form} can be rewritten as
\begin{align}
\vec{h}_t &= \mat{A}{\left(\vec{h}_{t-1}\right)} \cdot \vec{h}_{t-1} + \mat{P}^{(\textrm{i})} \cdot \vec{x}_t \label{eq:sup_ScalarRNN1}\\
\vec{y}_t &= \left\vert \mat{P}^{(\textrm{o})} \cdot \vec{h}_t \right\vert^2, \label{eq:sup_ScalarRNN2}
\end{align}
where we have defined $\mat{A}$ as the matrix appearing in Eq. (\ref{eq:matrix_form}).
The nonlinear dependence on $\vec{h}_{t-1}$ is defined by the nonlinear wave speed described above.

An absorbing region is introduced to approximate an open boundary condition \cite{oskooi_failure_2008}, corresponding to the grey region in Fig. \ref{fig:fig2}B. 
This region is defined by a damping coefficient, $b{(x,y)}$, which has a cubic dependence on the distance from the interior boundary of the layer.
The scalar wave equation with damping is defined by the inhomogeneous partial differential equation \cite{elmore_physics_2012}
\begin{equation}
    \frac{\partial^2{u}}{{\partial{t}}^2} +  2 b \cdot \frac{\partial{u}}{{\partial{t}}} = c^2 \cdot \nabla^2 u + f,
    \label{eq:wave_eq_damping}
\end{equation}
where $u$ is the unknown scalar field, $b$ is the damping coefficient. Here, we assume that $b$ can be spatially varying but is frequency-independent.
For a time step indexed by $t$, Eq. \ref{eq:wave_eq_damping} is discretized using \textit{centered} finite differences in time to give
\begin{align}
    \frac{u_{t+1} - 2u_t + u_{t-1}}{\Delta{t}^2} + 2b \frac{u_{t+1}-u_{t-1}}{2\Delta{t}} &= c^2 \nabla^2 u_t + f_t.
    \label{eq:wave_eq_damping_discretized}
\end{align}
From Eq. \ref{eq:wave_eq_damping_discretized}, we may form a recurrence relation in terms of $u_{t+1}$, which leads to the following update equation
\begin{align}
    \left( \frac{1}{\Delta{t}^2} + \frac{b}{\Delta{t}}\right) u_{t+1} - \frac{2}{\Delta{t}^2} u_t  + \left( \frac{1}{\Delta{t}^2} - \frac{b}{\Delta{t}} \right) u_{t-1} = c^2 \cdot \nabla^2 u_t + f_t \nonumber \\
    \left( \frac{1}{\Delta{t}^2} + \frac{b}{\Delta{t}}\right) u_{t+1} = \frac{2}{\Delta{t}^2} u_t  - \left( \frac{1}{\Delta{t}^2} - \frac{b}{\Delta{t}} \right) u_{t-1} + c^2 \cdot\nabla^2 u_t + f_t \nonumber \\
    u_{t+1} = \left( \frac{1}{\Delta{t}^2} + \frac{b}{\Delta{t}}\right)^{-1} \left[ \frac{2}{\Delta{t}^2} u_t - \left( \frac{1}{\Delta{t}^2} - \frac{b}{\Delta{t}} \right) u_{t-1} + c^2\cdot \nabla^2 u_t + f_t \right].
    \label{eq:tmp1}
\end{align}
Equation \ref{eq:tmp1} therefore represents the discretized update equation for the scalar wave equation with damping. In matrix form, we may express Eq. \ref{eq:tmp1} as
\begin{equation}
    \begin{bmatrix}
    u_{t+1} \\ u_t
    \end{bmatrix}
    = 
    \begin{bmatrix}
    \frac{2 + \Delta t^2\cdot c^2 \cdot \nabla^2}{1 + \Delta t\cdot b} 
    & \frac{-1 - \Delta t\cdot b}{1 + \Delta t\cdot b}  \\
    1 & 0
    \end{bmatrix}
    \cdot
    \begin{bmatrix}
    u_{t} \\ u_{t-1}
    \end{bmatrix}    
    +
    \Delta{t}^2 \cdot \begin{bmatrix}
     f_{t} \\ 0
    \end{bmatrix},
    \label{eq:update_damped}
\end{equation}
which also has the same form as in Eq. \ref{eq:eq5} and \ref{eq:eq6} of the main text.

\subsection{Realistic physical platforms and nonlinearities \label{sec:nonlinearities}}

The results presented in the main text were obtained using a very general physical model for the wave dynamics, where the field is assumed to be scalar and the hidden state to hidden state nonlinearity originates from an intensity-dependent wave speed.
In this section, we discuss in more detail how the scalar wave results can be translated into various practical platforms, in both optics and acoustics, using various materials as well as different forms of nonlinearity.
The key experimental considerations for a practical realization of the analog wave RNN are (1) achieving a compact (wavelength-scale) physical footprint, (2) a physical medium into which the physical weights can be patterned, and (3) a nonlinear material response which can be achieved at reasonable signal energies, without requiring high power sources.
We emphasize that the introduction of a nonlinearity into the wave dynamics is critical to realizing the complex information processing capabilities akin to a conventional RNN, allowing the wave dynamics to extend beyond what is achievable with linear time-invariant system theory.

\subsubsection*{Optics}

The most straightforward realization of the nonlinear wave speed in optical platforms is using Kerr nonlinearity.
Silicon (Si) and the family of chalcogenide glasses (e.g. As$_2$S$_3$) are two widely used nonlinear optical materials for integrated platforms, with chalcogenide having one of the highest damage thresholds.
Such a high threshold allows for processing of sub-picosecond laser pulses with peak powers on the order of 10-50 MW \cite{lamont2008supercontinuum}.
A longer pulse duration in such nonlinear materials will lead to irreversible damage and imposes an upper bound on the pulse length which can be processed by the analog RNN.
Such an ultra-fast optical analog RNN with Kerr nonlinearity may be useful in a number of scientifically-relevant applications, such as diagnostics and processing for ultrafast pulses in nonlinear spectroscopy and X-ray free electron lasers \cite{hartmann2018attosecond}.
In such applications, the integration of conventional electronic processors is extremely challenging due to the arrival rate of information as well as environmental factors.
Additionally, a potential advantage of sub-ps optical analog RNNs is that the training process for the RNN will not need to access regions of the device parameter space with ultra-narrowband (high-Q) spectral features because the relative signal bandwidths of sub-ps pulses are large. 
In contrast, an optical analog RNN for processing optical carriers broadened by GHz-rate electro-optic modulators will likely be required to operate in the parameter space associated with sub-GHz, or even sub-MHz, spectral features in order to effectively learn features found in such narrowband signals.

\begin{figure*}
  \centering
  \includegraphics{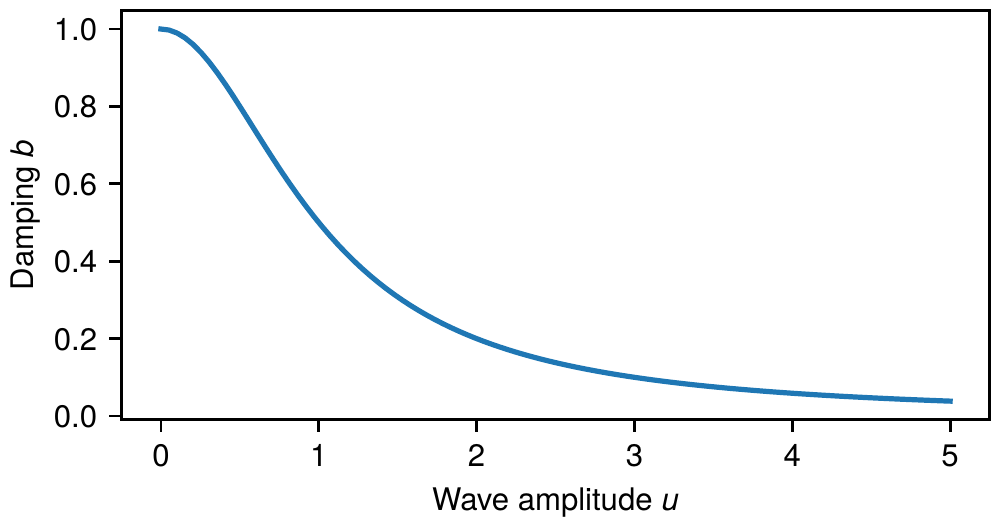}
  \caption{Saturable absorption response, for the parameters $b_0 = 1.0$ and $u_{\text{th}} = 1.0$ in Eq. \ref{eq:saturable_absorption}, indicating the nonlinear dependence of the damping parameter, $b$ on the wave amplitude, $u$.}
  \label{fig:sat_abs}
\end{figure*}

An alternative optical nonlinearity which could be used to construct an analog RNN is saturable absorption. Such a nonlinear response consists of an intensity-dependent absorption/damping, which is mathematically defined as
\begin{equation}
    b{\left( u \right)} = \frac{b_0}{1+\left(\frac{u}{u_{\text{th}}}\right)^2}, \label{eq:saturable_absorption}
\end{equation}
where $b_0$ and $u_{\text{th}}$ are the saturable absorption strength and threshold, respectively.
An example of the saturable absorption response is plotted in Fig. \ref{fig:sat_abs}.
One potential realization of this effect involves the patterning of graphene or other absorptive 2D materials on top of the linear optical circuit etched into a dielectric such as silicon.
An advantage of saturable absorption over the Kerr effect is that the resulting nonlinear response can be observed at input powers on the mW scale \cite{jiang2018low}, making saturable absorption a promising candidate for use in optical analog RNNs.
On the other hand, the main disadvantage of saturable absorption is that it inevitably introduces power loss into the system, which could potentially degrade the signal to noise ratio (SNR) at the detectors in a large-scale analog RNN.
Because the complexity of the analog RNN is directly related to its physical footprint, this may limit the expressive capability of an RNN using this form of nonlinearity.
However, if we assume a saturable absorption threshold intensity of 0.5 MW/cm$^2$ for graphene \cite{jiang2018low}, a structure that is 10$\lambda$ wide [the $y$-extent of the structure in Fig. \ref{fig:fig2}(D)] at an operating wavelength of $\lambda=1550$ nm, and an out-of-plane thickness of 1 $\mu$m, this situation would require an input power of 77 mW to achieve the intensity threshold across the input facet of the analog RNN.
This nonlinear threshold is orders of magnitude lower than that of the Kerr effect, making saturable absorption an appealing nonlinearity for the analog RNN.

\begin{figure*}
  \centering
  \includegraphics{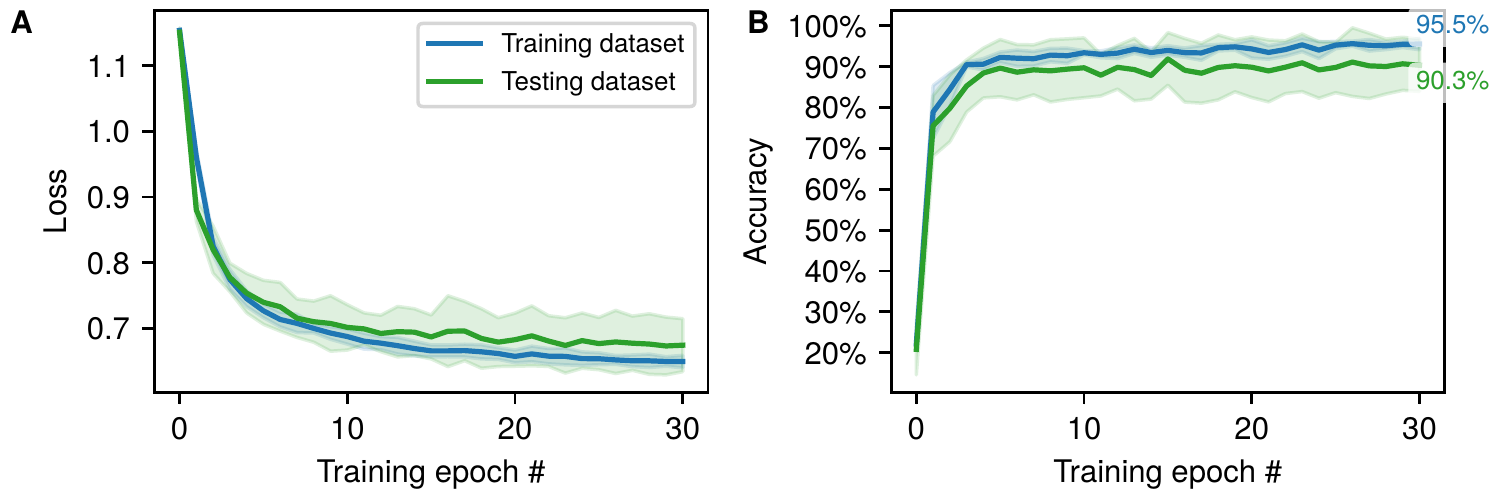}
  \caption{Cross validated training results for an RNN with a saturable absorption nonlinearity. The mean (sold line) and standard deviation (shaded region) of the (\textbf{A}) cross entropy loss and (\textbf{B}) prediction accuracy over 30 training epochs and 5 folds of the dataset, which consists of a total of 279 total vowel samples of male and female speakers. The parameters used for the saturable absorption (Eq. \ref{eq:saturable_absorption}) in these results are $b_0 = 0.1$ and $u_{\text{th}} = 2.5\times 10^{-4}$, while a batch size of 15 samples was used during training.}
  \label{fig:sat_abs_results}
\end{figure*}

As an example, we numerically demonstrate a version of the wave RNN with saturable absorption. 
The training results for this system are shown in Fig. \ref{fig:sat_abs_results}(A) and (B), where we observe that this form of nonlinearity can also perform well on the vowel classification task from the main text, achieving a training and testing classification accuracy of $95.5\% \pm 1.4 \%$ and $90.3\% \pm 6.4\%$, respectively.
These accuracies are comparable to those of the analog RNN with a nonlinear wave speed as shown in Fig. \ref{fig:fig3} of the main text.
We note however, that the saturable absorption nonlinearity results in a larger variance in the accuracy on the testing set over the 5 cross validated training runs.
The increase in variance could be due to variations in the peak signal amplitudes of the various vowel recordings, even though we normalize all recordings to have equal time-integrated power.
Essentially, some vowel samples may not be ``loud'' enough to overcome the damping and, thus, are nearly completely absorbed before they reach the detectors.

In terms of the device's physical footprint, although we have considered a 2D simulation, the physics observed in such systems translates well to planar integrated optical circuits. In such 3D devices, confinement of light in the out-of-plane direction can be achieved using index confinement. We note that similar planar optical circuits produced via inverse-design techniques have been previously demonstrated experimentally \cite{molesky2018inverse}.

\subsubsection*{Acoustics}

An acoustic or elastic wave implementation of the wave RNN has several advantages over optical platforms in terms of the availability of off-the-shelf components and nonlinear material responses, which can be much stronger. 
Moreover, the typical operating frequencies used in acoustic signal processing are also orders of magnitude smaller than typical optical carrier frequencies.
This, in turn, naturally leads to much larger relative signal bandwidths and eliminates the requirement of realizing ultra-narrowband spectral features in the parameter space of the analog RNN.
An acoustic version of the analog RNN could utilize off-the-shelf freespace ultrasonic transducers and receivers in conjunction with a 3D-printed or laser-cut polymer.
Many polymers can exhibit a \textit{slower} sound speed than in air \cite{ba_soft_2017}, as is the case in the system demonstrated in the main text. 
However, we emphasize that a lower sound speed in the printed material is \textit{not} a fundamental requirement for the RNN.
In principle, so-called \textit{hard} material sidewalls could also be utilized, similarly to the structures of previously demonstrated acoustic metamaterials \cite{christiansen_experimental_2016}, where $c > 2000$ m/s.
It's also worth pointing out that sub-wavelength 3D-printed inclusions could be treated via an effective medium theory in order to achieve a range of sound speeds.
This approach would offer interesting opportunities for non-binarized implementations of the analog RNN.

In terms of a nonlinear material response, many fluids, particularly those with embedded gas bubbles (e.g. carbonated water), exhibit a strongly nonlinear response.
This effect is captured through a Taylor expansion of the fluid's equation of state, which defines the relationship between pressure, density, and entropy \cite{rossing2007springer}.
A common approach for modeling such effects is using the nonlinear acoustic Westervelt equation \cite{pinton_heterogeneous_2009}, which includes several terms in addition to the ones in the linear wave dynamics defined by Eq. \ref{eq:wave_eq_damping}. 
Typically, nonlinear fluids exhibit a second-order nonlinear response \cite{rossing2007springer}, where $c{\left(u\right)} \sim u$, which is different from the third-order nonlinearity we use in the main text, but is qualitatively similar.
The Westervelt equation also includes a term that accounts for thermoviscous damping, which introduces a frequency-dependent attenuation.
One potential route towards including such nonlinearities into an experimental realization of the RNN would be to infiltrate a 3D-printed linear structure with a highly nonlinear fluid.
Although, this approach would be different from the system considered in the main text, in terms of which material includes a nonlinear response, it would represent a practical realization of the analog RNN.
A non-freespace platform for implementing the RNN would be to use elastic Lamb waves or surface waves on patterned slabs.
The features of such systems could be defined lithographically and nonlinearities could again be achieved by infiltrating the patterned linear material with a nonlinear fluid.

In summary, this section shows that there are several realistic pathways for implementing the wave-based RNN described in the main text, including the nonlinearities.

\subsection{Input and output connection matrices \label{sec:io}}

In this section we discuss, in detail, the linear operators, $\mat{P}^{(\textrm{i})}$ and $\mat{P}^{(\textrm{o})}$, that define the injection and measurement locations within the domain of the wave equation. We start from the vectors $\vec{u}_t$ and $\vec{f}_t$, that are discretized and flattened vectors from the field distribution $u_t$ and $f_t$.
Then, we define the linear operators, $\mat{M}^{(\textrm{i})}$ and $\mat{M}^{(\textrm{o})}$, each column of which define the respective spatial distributions of the injection and measurement points in this flattened basis.
With this, we can write the injection of the input vector, $\vec{x}_t$ as a matrix-vector multiplication
\begin{equation}
    \Delta t^2 \vec{f}_t \equiv \mat{M}^{(\textrm{i})} \cdot \vec{x}_t.
\end{equation}

Similarly, as the output of the RNN at each time step is given by an intensity measurement of the scalar fields, we may express this in terms of the flattened scalar field as
\begin{equation}
    \vec{y}_t = \mat{M}^{(\textrm{o})}{}^T \cdot {\vec{u}_t}^2.
\end{equation}
As the wave equation \textit{hidden state}, $\vec{h}_t$ is defined as the concatenation of $\vec{u}_t$ and $\vec{u}_{t-1}$, we define the following matrices for convenience, as they only act on the $\vec{u}_t$ portion of $\vec{h}_t$
\begin{align}
    \mat{P}^{(\textrm{i})} &\equiv \begin{bmatrix} \mat{M}^{(\textrm{i})} \\ \mat{0} \end{bmatrix} \\ 
    \mat{P}^{(\textrm{o})} &\equiv [\mat{M}^{(\textrm{o})}{}^T,~\mat{0}],
\end{align}
where $\mat{0}$ is a matrix of all zeros.
These matrices are used in the injection and measurement stages of the scalar wave update equations of the main text and thus serve a similar role to the $\mat{W}^{(x)}$ and $\mat{W}^{(y)}$ matrices of the traditional RNN in Eqs. \ref{eq:eq1} and \ref{eq:eq2}.  However, unlike $\mat{W}^{(x)}$ and $\mat{W}^{(y)}$, these matrices are fixed and not trainable parameters. In our numerical implementation, the operation of $\nabla^2$ on a spatially discretized wave field, $u_t$ is computed using the convolution operation, defined mathematically as
\begin{equation}
    \nabla^2 u_t =
    \frac{1}{h^2}
    \begin{bmatrix}
    0 &  1 & 0 \\
    1 & -4 & 1 \\
    0 &  1 & 0
    \end{bmatrix}
    *
    u_t,
    \label{eq:convolution}
\end{equation}
where $h$ is the step size of the spatial grid.

\subsection{Comparison of wave RNN and conventional RNN \label{sec:compare}}

In this section, we compare the performance of the wave RNN to that of a conventional RNN, as defined by Eq. \ref{eq:eq1} and Eq. \ref{eq:eq2}. 
In the conventional RNN, the number of trainable parameters is determined by the size of the hidden state, $N_h$, where the model is parameterized by three matrices $\mat{W}^{(x)}, \mat{W}^{(h)}$, and $\mat{W}^{(y)}$ of size $\left[ N_h \times 1 \right]$, $\left[ N_h \times N_h \right]$, and $\left[ 3 \times N_h \right]$, respectively. 
We selected $N_h = 70$ and $N_h=100$, which correspond to a total number of free parameters in the RNN of 5250 and 10500, respectively. 
This RNN model was implemented and trained using the \texttt{pytorch} framework.  
In Table \ref{tab:final_table} we compare the final prediction accuracy of the conventional RNN on the vowel recognition task to the wave RNN, where we observe that the conventional RNN can achieve a performance comparable to that of the wave RNN.
However, the conventional RNN is very sensitive to the total number of trainable parameters.  
For a similar number of trainable parameters to that of the wave RNN, the conventional RNN achieves approximately 6\% lower classification accuracy.  
However, when the number of free parameters is increased to about twice that of the wave RNN, the accuracy is higher by approximately 3 \%. 
We note that it may be possible to achieve higher accuracy in more advanced recurrent models such as the long short-term memory (LSTM) \cite{hochreiter1997long} or gated recurrent unit (GRU) \cite{chung2014empirical} architectures.
However, a detailed exploration of these models is outside the scope of the current study. 

The conventional RNN and the wave RNN have a number of qualitative differences which we now discuss in more detail.
First, in the conventional RNN, the trainable parameters are given by the elements of the weight matrices.  
In the case of the wave RNN, we choose to use the wave speed, $c(x,y,z)$, defined on a discretized grid to define the set of trainable parameters, because a specific distribution of $c$ can be physically implemented after the training process.  
Moreover, while the free parameters of the conventional RNN define a matrix which is multiplied by the input, output, and hidden state vectors, in the wave RNN, the free parameters are multiplied element-wise with the hidden state, which limits the influence of each individual parameter over the full dynamics of information within the hidden state. 
For a given amount of expressive power, the size of the hidden state in the wave equation must arguably be much larger than that of the conventional RNN.  
The reason for this is that the amount of information which can be encoded into the spatial distribution of $u_t$ is constrained by the diffraction limit for waves.  
Thus, it follows that a single element from the hidden state of a conventional RNN may be analogous to several grid cells in the scalar wave equation. 
Furthermore, the discretized wave equation update matrix, $A$, is sparse and only contains non-zero values around its main diagonal, which physically corresponds to a neighbor-to-neighbor coupling between spatial grid cells (through the Laplacian operator).  
Due to this form of coupling, information in a given cell of $u_t$ will take many time steps before interacting with information stored in distant cells, as determined by the wave velocity and the physical distance between them.  
The presence of this form of causality practically means that one must wait longer for a full `mixing' of information between cells in the domain, suggesting that in the our numerical simulations, a larger number of time steps may be needed as compared to the typical RNN. 

Finally, the form of nonlinearity used in the wave RNN is conceptually distinct from that used in the conventional RNN, which involves the application of the nonlinear function, $\sigma^{(h)}(\cdot)$, as in Eq. \ref{eq:eq1}.  
In the wave RNN, nonlinearity is provided by making the wave velocity, $c$, or damping, $b$, to be dependent on the instantaneous wave intensity ${u_t}^2$, i.e. $c = c{\left({u_t}^2\right)}$, or $b = b{\left({u_t}^2\right)}$.  
With this addition, the update matrix of Eq. \ref{eq:eq5}, $\mat{A} = \mat{A}(\vec{h}_{t-1})$, becomes a function of the solution at that time step, making the dynamics nonlinear.
Nonlinearity is introduced into the output of the wave system ($\vec{y}_t$) through a measurement the wave intensity, which involves a squaring operation.  
The practical realization of material nonlinearities is discussed in detail in the supplementary materials section \ref{sec:nonlinearities}.

\begin{table*}[t]
  \centering
  \setlength{\tabcolsep}{6pt}
    \renewcommand{\arraystretch}{1.25}
  \begin{tabular}{llccccc}
    \hline\hline
    \textbf{Model} & \textbf{Nonlinearity} & \textbf{\# parameters} & \multicolumn{2}{c}{\textbf{Accuracy}} \\
    & & & \textbf{Training} & \textbf{Testing} \\
    \hline
    \textbf{Wave Equation} & linear wave speed & 4200 & 93.1\% & 86.6\% \\
      & nonlinear wave speed & 4200 & 92.6\% & 86.3\% \\
      & saturable damping & 4200 & 95.5\% & 90.3\% \\
    \hline
    \textbf{Conventional RNN} & linear & 5250 & 78.8\% & 79.4\% \\
    & leaky ReLU & 5250 & 82.6\% & 80.2\% \\
    & linear & 10500  & 88.9\% & 88.2\% \\
    & leaky ReLU     & 10500 & 89.4\% & 89.4\% \\
    \hline\hline
  \end{tabular}
  \caption{Comparison of scalar wave model and conventional RNN on vowel recognition task. The saturable damping nonlinearity is described in detail in the next section.}
  \label{tab:final_table}
\end{table*}

\subsection{Binarization of the wave speed distribution \label{sec:binarization}}

In this section we discuss how we create realistic distributions of material with a binarized $c(x,y)$ distribution using filtering and projection schemes during our optimization. Rather than updating the wave speed distribution directly, we instead update a design density $\rho(x,y)$, which describes the \textit{density} of material in each pixel.  To create a structure with larger feature sizes, a low pass spatial filter can be applied to $\rho(x,y)$ to created a filtered density,
\begin{equation}
\tilde{\rho}(x,y) = 
\begin{bmatrix}
1/9 & 1/9 & 1/9 \\
1/9 & 1/9 & 1/9 \\
1/9 & 1/9 & 1/9
\end{bmatrix}
*
\rho(x,y).
\end{equation}

For binarization of the structure, a projection scheme is used to recreate the final wave speed from the filtered density.  We define $\bar{\rho}(x,y)$ as the projected density, which is created from $\tilde{\rho}(x,y)$ as
\begin{equation}
\bar{\rho}_i = \frac{\tanh{\left( \beta \eta \right)} + \tanh{\left( \beta \left[ \tilde{\rho}_i - \eta \right] \right)}}{\tanh{\left( \beta \eta  \right)} + \tanh{\left( \beta \left[ 1 - \eta \right] \right)}}.
\end{equation}
Here, $\eta$ is a parameter between a value of 0 and a value of 1 that controls the mid-point of the projection, typically 0.5, and $\beta$ controls the strength of the projection, typically around 100. The distribution of $\bar{\rho}$ varies between 0 and 1.
Finally, the wave speed can be determined from $\bar{\rho}$ as
\begin{equation}
c(x,y) = (c_1 - c_0)\bar{\rho}(x,y) + c_0,
\end{equation}
where $c_0$ and $c_1$ are the background and optimized material wave speed, respectively.

\end{document}